\documentstyle[12pt]{article}
\begin{document}

\begin{titlepage}
\begin{flushright}
       {\bf UK/94-04}  \\
 December 1994      \\
      hep-lat/9501007
\end{flushright}
\begin{center}
\vspace{1cm}

{\bf {\LARGE Finite $ma$ corrections for sea quark matrix elements on the
lattice}}

\vspace{1cm}

 {\bf J.-F. Laga\"e and K.-F. Liu}
 \\ [0.5em]
 {\it Dept. of Physics and Astronomy  \\
  Univ. of Kentucky, Lexington, KY 40506}

\end{center}

\vspace{1cm}

\begin{abstract}

We compute the $ma$ dependence of lattice renormalization factors for
sea quark matrix elements. The results differ from the $(1+ma)$ correction
factor commonly used for valence quarks and connected current insertions.
We find that for sea quarks, the correction factors are in general larger
and depend strongly on the Lorentz structure of the current under
consideration. Results are presented both for the Wilson action and for
the 2-link improved action of Hamber and Wu. Phenomenological implications
are also briefly discussed in two examples.

\vskip\baselineskip

\end{abstract}

\vfill

\end{titlepage}

\section{Introduction}

Many interesting problems in hadronic physics (e.g. the proton spin crisis,
the U(1) problem, the strange quark content of the nucleon, ...) require the
computation of a quark current inside of a disconnected\footnote
{Here the terms connected or disconnected refer only to the
topological classification of the quark lines (regardless of the presence
of gluonic lines).} fermionic loop. Matrix elements in the nucleon for example
generally involve two types of contribution associated with the connected
(fig.1.a) or disconnected (fig.1.b) insertion of the quark current. Similarly,
the propagation of flavor-singlet mesons involves, besides the usual connected
contribution (fig.1.c), a disconnected contribution (fig.1.d). Up to recently,
only the contributions associated with fig.1.a and c (which we will call
``valence'' for short\footnote
{Actually, the connected insertions contain both valence quarks and cloud
quarks and antiquarks \cite{Liu94}.}) were considered in lattice simulations.
Contributions
from fig.1.b and d (which we will call ``sea'' for short), on the other hand,
were considered too difficult to compute. Recent developments in the
technology of lattice gauge theory have however made their computation
feasible. Among the techniques currently being used are the $Z_2$ noise method
\cite{Dong94,Lat94a}
and the non-gauge-fixed volume source technique \cite{Kura94,Fuku94}.
Progress is fast in this
area and numerical simulations based on these methods are beginning to
produce statistically significant results. It therefore becomes essential
to compute the renormalization factor associated with sea quark
matrix elements in order to turn the raw numbers from Monte-Carlo simulations
into physical quantities.
The lattice renormalization of bilinear quark operators (in the limit
$ ma \rightarrow0 $) was considered a long time ago \cite{Marti83},
but the finite $ma$ corrections were only
introduced more recently \cite{Lepa92}
along with the tadpole improvement \cite{Lepa93}.
Here we would like to get a closer look at the finite $ma$ corrections in
the particular case of sea quark matrix elements. We expect this to be
important for the problem at hand. One would like for example to be able to
measure various strange quark matrix elements in the nucleon. In current
simulations, $ma$ is about $0.1$ for a strange quark and therefore finite
$ma$ corrections might be significant in this case, and indeed they are as
we will see below.
At the heart of the present study is the realization that the finite $ma$
correction factor commonly used for valence quarks and connected current
insertions is not adequate for the case of sea quarks or disconnected current
insertions.
In section 2, we briefly review the usual approach and
discuss its limitations. We also make some general comments which will be
useful in understanding the results presented later. In section 3, we
motivate and introduce the method that we use to compute the finite
$ma$ correction factors for sea quarks. The method in itself is rather
straightforward and mostly consists in computing the perturbative response
of a quark current to an external gauge field both on the lattice and in
the continuum. By comparing the two results, the appropriate correction
factor is then extracted. We have carried out the procedure for the scalar,
pseudoscalar and axial currents which are of most direct phenomenological
relevance. The computation of the relevant Feynman diagrams in the continuum
is presented in section 4, whereas the lattice results are to be found in
section 5. The finite $ma$ correction factors are then computed and the
results are summarized in table 1 for the Wilson action and in table 2 for
the 2-link improved action of Hamber and Wu \cite{Wu84}. The application of
these correction factors to Monte-Carlo data is discussed in section 6 for
two examples (the computation of the mass of the $\eta^\prime$ and the
determination of the strange quark contribution to the mass of the nucleon).
In section 7, we add some final remarks and briefly discuss further
developments. Our conventions and Feynman rules are presented in the appendix.

\section{Wilson fermions at finite $ma$}

It has been customary to estimate the finite $ma$ corrections by comparing
the lattice and continuum propagators of a {\it free quark} at ${\vec p} = 0 $
\cite{Lepa92}.
The result of this computation gives a $(1+ma)$ finite renormalization factor
for quark bilinears, where $ma$ is the mass appearing in the Lagrangian or
more generally the mass obtained after tadpole improvement:
   \begin{equation}
    ma= {1 \over 2\tilde\kappa} - 4
   \quad\quad \mbox{where} \quad\quad
   \tilde\kappa=\kappa U_0 \simeq {\kappa \over 8\kappa_c}
   \end{equation}
The assumptions going into this calculation appear to be justified for
valence quarks. If the quark is heavy, its 3-momentum is small and it is
``almost on mass-shell'', whereas when $ma$ goes to $0$ the correction factor
goes back to $1$. So the approximation is reasonable for a quark
propagating forward in time inside a meson or baryon at rest (although
there seems to be room for improvement at small and intermediate values
of $ma$).
For sea quark matrix elements, however, the situation is completely different.
Configuration by configuration, one is computing a closed quark loop in a
background gauge field (see fig.1.b and d). This involves an integration
over the 4-momentum of the quark running around the loop. Therefore, the
kinematical restrictions (i.e. low 3-momentum and proximity to the mass-shell)
which were justified for valence quarks no longer apply for sea quarks.
A more general method, avoiding particular kinematical assumptions, will
thus be necessary to compute the finite $ma$ corrections in this case.

Before introducing our technique for dealing with sea quarks in section 3,
we would like to make here some qualitative remarks on the nature of the
problem. As is well known, the $r=0$ theory (naive action) contains
16 fermions instead of 1 and the role of the Wilson term is to
lift the degeneracy between the various species. In the limit
$ ma \rightarrow 0 $, it does so no matter what the value of $r (\neq 0)$ is.
However when $ma$ is non-zero, the effectiveness of the Wilson term in
lifting the degeneracy becomes more limited and r dependent.
In this sense, at finite ma, one is dealing with an effective theory
with a more complicated flavor content. There are 16 species with masses
\cite{Smit81}:
$$\begin{array}{ccccc}
  \quad ma \quad & \quad ma+2r \quad & \quad ma+4r \quad & \quad ma+6r \quad &
  \quad ma+8r \quad \\
  \{1\} &  \{4\} &  \{6\}  &  \{4\}  &  \{1\}
\end{array}$$
where the degeneracy of each level is indicated between braces.
It is clear that when $ma$ becomes of the same order as $r$, the doublers
are no longer much heavier than the ``fundamental'' fermion. This causes
strong lattice artifacts, in the computation of closed
quark loops and requires the use of correction factors in order to map
the effective ``multiflavor'' theory onto QCD. It is precisely the
computation of these finite renormalizations that we will consider in
this paper. We will see that they depend on the Lorentz structure of
the current under consideration. This is to be expected. Take {\it naive}
fermions for example, matrix elements of the scalar current in a smooth
external gluonic field would be 16 times what
they should be in QCD, whereas matrix elements of the axial current would
be 0 in the chiral limit because the chiralities of the 16 doublers add
to $0$ \cite{Smit81}.
As we will see below similar trends subsist for Wilson fermions at finite
$ma$: matrix elements of the scalar current are overestimated
whereas those of the axial current are underestimated. That is,
at finite $ma$, the Wilson term only takes care of the doubling
problem partially and this affects the measurements of sea quark
matrix elements.

\section{Renormalization of sea quark matrix elements}

Since configuration by configuration, the computation of a sea quark matrix
element reduces to the computation of an ``induced quark current'' in a given
background gauge field, we are led to study the ``response functions'' to an
external gluonic field. By computing the matrix element of some current in
a given background gauge field on the lattice (at finite $ma$) and comparing
with the induced current in the ``same'' external field in the continuum, we
can estimate the lattice artifacts at this particular value of $ma$ and
correct for them.
A particularly interesting example was considered by Smit and Vink several
years ago \cite{Smit87}. They investigated the case of the flavor-singlet
pseudoscalar current where a topological relation gives an exact expression
for the response function. Indeed, by making use of the anomalous Ward identity
   \begin{equation}
   \partial_\mu j_{\mu 5} -2 m j_5 = 2 n_f q
   \end{equation}
where q is the topological charge density defined as
   \begin{equation}
   q = {1 \over 32\pi^2} \epsilon_{\mu\nu\rho\sigma}
       Tr( G_{\mu\nu} G_{\rho\sigma} )
   \end{equation}
and integrating over 4-volume, one obtains a direct relation between the
zero momentum matrix element of $j_5$ ($Q_5 = \int d^4x j_5$)
and the topological charge ($Q_t = \int d^4x q$)
of the gauge field configuration:
   \begin{equation}
   m Q_5 = - n_f Q_t
   \end{equation}
It is then sufficient to take a smooth background
field configuration of given topological charge on the lattice, to
measure the matrix element of $Q_5$ at various values of $ma$ and to
compare with (4) to be able to deduce the appropriate $ma$ dependent
correction factor. We will call this factor $\kappa_P(ma)$
[P for pseudoscalar]. Smit and Vink then went on to
compute the topological susceptibility through correlations of
$<j_5(x)j_5(0)>$. Taking into account the correction factor,
one obtains in terms of the lattice currents:
   \begin{equation}
   \chi = <Q_t^2>/V = \kappa_P^2(ma) \int d^4x <j_5(x)j_5(0)> m^2 / n_f^2
   \end{equation}
Strictly speaking, one has to be careful at this point, since the correction
factor could be gauge configuration dependent, in which case normalizing on
a single configuration, as was done above, would not be adequate. However, for
the present computation, a consistency check can be used to show that finite
$ma$ artifacts are to a good approximation configuration independent. We know
that the topological susceptibility (being a pure glue quantity in the
quenched approximation) must be independent of the mass of the quarks.
An evaluation of (5) on a statistical sample of gauge configurations would
however not satisfy this constraint if $\kappa_P(ma)$ wasn't the appropriate
correction factor for every ``important'' configuration. The fact that Smit
and Vink indeed obtain for (5) a mass independent result in their Monte-Carlo
simulation therefore can be used as an indication that a single, configuration
independent, correction factor [$\kappa_P(ma)$] is sufficient.

Now we would like to generalize the computation of finite $ma$ corrections to
other currents. The pseudoscalar current was special because of the existence
of an exact ``topological'' relation (4); other cases won't be that
straightforward. However, if we assume that the $ma$ dependence of the
correction factor is the same for all configurations (as seems to be the case
from the above discussion), any background gauge field can be used for its
evaluation. In particular, we can consider very weak external fields, so
that perturbation theory becomes a reliable tool, and compute in this manner
the relation between lattice and continuum theory.
This is the strategy that we will adopt here. It turns out that the lowest
order perturbative diagrams for the expectation value of a quark current
$\bar\psi\Gamma\psi$ in an external field will be triangle diagrams
like the one represented in fig.2.a .
In this paper, we will restrict our attention to matrix
elements of zero momentum currents (so that $q=-p$ in fig.2).
This simplifies the computations and is sufficient for the applications
currently envisaged (``forward'' matrix elements in the nucleon, for example).

Let's consider the scalar current as an example ($\Gamma = 1$ in fig.2) .
It is well known that in this case, the continuum computation of the triangle
diagram, gives a relation between the matrix element of the scalar current and
the square of the external field strength of momentum p (the zero momentum
relation has been abundantly used in the context of QCD sum rules \cite{SVZ}):
   \begin{equation}
   m<\bar{\psi}\psi> = C( {p^2 \over m^2} ) F^2_{\mu\nu}(p)
   \end{equation}
where $C$ is a dimensionless function of $p^2/m^2$. Similarly, one can
compute the triangle diagram on the lattice (see section 5 for details)
and obtain the corresponding relation:
   \begin{equation}
   m<\bar{\psi}\psi> = L(pa,ma) F^2_{\mu\nu}(p)
   \end{equation}
the only difference being that the lattice dimensionless function $L$
can depend separately on $pa$ and $ma$. The functions $L$ and $C$
contain all the information we need, they tell us how the lattice
and continuum scalar currents react to a given external field.
By comparing the two, we can deduce which correction factor need to be
introduced in lattice measurements of the scalar current. We will mostly
consider the case p=0 (corresponding to smooth background gauge fields).
Then one obtains the finite ($ma$ dependent)
renormalization factor for the scalar current:
   \begin{equation}
   \kappa_S(ma) = C(0)/L(0,ma)
   \end{equation}
Of course, one can check that $L(pa,ma)$ goes back to $C(p^2/m^2)$ when
both $pa$ and $ma$ are small so that $\kappa_S(ma) \rightarrow 1$ when
$ma \rightarrow 0$. If the numerical simulation is dominated by
smooth gauge fields, then $\kappa_S(ma)$ is all that we need. However,
this is not in general the case and it is therefore important to look
also at the situation at finite $pa$. To investigate this momentum
dependence, we will present below comparative plots of $L(pa,ma)$
and $C((pa)^2/(ma)^2)$ at fixed values of $ma$.
The perturbative method presented here for the scalar current
generalizes, mutatis mutandis, to other currents. One can for example
compute the finite $ma$ correction factor for the pseudoscalar current
and then compare with the topological charge method presented earlier.
The two results agree provided that the ``instanton'' field is weak
enough.

A general remark is in order here. In the case of valence quarks
\cite{Lepa92,Lepa93}, it is assumed that the $ma$ and $\beta$(or $\alpha_s$)
dependence of the renormalization factor factorize:
  \begin{equation}
  Z={2\tilde\kappa} (1+ma) [1+O(\alpha_V)]
  \end{equation}
We make the same factorization hypothesis here. This is what allows the
computation of the $ma$ dependence at lowest non-trivial order in
perturbation theory.  The actual order at which the computation is done
however differs for valence and sea quarks. In the valence case, the
ma dependence is computed for a free quark. For the correlation between
two closed quark loops (fig.1.d), on the other hand, the lowest non-trivial
order is $\alpha_s^4$ [or $\alpha_s^2$ for a single quark loop in a background
field]. This is why for sea quarks, we extract the $ma$ dependence from
the triangle diagram (fig.2). So that in summary, given the factorization
hypothesis,
what we compute here for sea quarks is the factor replacing the $(1+ma)$
factor of valence quarks; the rest of the renormalization factor
($[1+O(\alpha_V)]$ and tadpole corrections) remains unchanged.
Another way to look at this factorization for sea quarks would be to view
the gauge field configurations as a smooth background field with additional
short wavelength fluctuations. In this picture, the $ma$ dependent factor
corrects the response to the smooth background field and the perturbative
factor accounts for the short wavelength fluctuations.

\section{Response to continuum background fields}

The continuum computation of diagrams like the one presented in fig.2.a is
rather straigthforward. The Feynman rules that we have used are presented
in the appendix. These diagrams are finite and the only subtelty in their
computation is the introduction of a Pauli-Villars regulator (which simply
consists in substracting the same fermionic loop for an infinitely heavy quark)
to ensure gauge invariance. We have done the computation for the
axial, scalar and pseudoscalar currents ($\Gamma=\gamma_\alpha\gamma_5,
\gamma_5$ and $I$ respectively) which are of most direct phenomenological
relevance. As mentioned above, we will restrict our attention to zero
momentum currents so that $q=-p$ in fig.2 .
For the pseudoscalar current, however, this would lead to a vanishing
result and we therefore give the general formula in this case.
Finally, we should mention that when quoting our final
results, we consider not only the diagram of fig.2.a, but the sum of this
diagram and the one with crossed photon legs [fig.2.b]
(We do not include the external gauge fields and couplings in our
expressions, since they are just common factors which can be brought
back at the end). This gives us respectively,
for the axial current ($q=-p$):
   \begin{equation}
   A_{\alpha\mu\nu}(p,-p) =
   -{1 \over \pi^2} \epsilon_{\mu\nu\alpha\beta} p_{\beta} A({p^2 \over m^2})
   \end{equation}
for the scalar current ($q=-p$):
   \begin{equation}
   S_{\mu\nu}(p,-p) =
   {1 \over 6\pi^2m} (\delta_{\mu\nu}p^2-p_{\mu}p_{\nu}) S({p^2 \over m^2})
   \end{equation}
and for the pseudoscalar current:
   \begin{equation}
   P_{\mu\nu}(p,q) =
   {1 \over 8\pi^2m} \epsilon_{\mu\nu\alpha\beta} q_{\alpha} p_{\beta}
   P( {p^2 \over m^2} , {q^2 \over m^2} , {p \cdot q \over m^2} )
   \end{equation}
where the functions A,S,P are given by:
   \begin{equation}
   A({p^2 \over m^2}) = {1 \over 2} \left\{
   1 - {m^2 \over p^2} {1 \over \sqrt{\quad} }
       ln({ \sqrt{\quad} + 1/2 \over \sqrt{\quad} - 1/2}) \right\}
   \end{equation}
   \begin{equation}
   S({p^2 \over m^2}) = 12 ({m^2 \over p^2}) A ({p^2 \over m^2})
   \end{equation}
   \begin{equation}
   P( {p^2 \over m^2} , {q^2 \over m^2} , {p\cdot q \over m^2} )
   = \int_0^1 dx \int_0^{1-x} dy
   { 2m^2 \over q^2x(1-x)+p^2y(1-y)+2p\cdot q xy + m^2 }
   \end{equation}
with
   \begin{equation}
   \sqrt{\quad} = \sqrt{ {1 \over 4} + {m^2 \over p^2} }
   \end{equation}
Note that all these expressions lead to gauge invariant results (even for
the axial current since $q=-p$).
Finally, by using the asymptotic expansion of the functions $A$, $S$ and $P$,
one obtains the low momentum limits of (10), (11) and (12):
   \begin{equation}
   A_{\alpha\mu\nu}(p,-p) \stackrel{p\rightarrow0}{\longrightarrow}
-{1 \over 12\pi^2} \epsilon_{\mu\nu\alpha\beta} p_{\beta} {p^2 \over m^2} \\
   \end{equation}
   \begin{equation}
   S_{\mu\nu}(p,-p) \stackrel{p\rightarrow0}{\longrightarrow}
 {1 \over 6\pi^2m} (\delta_{\mu\nu}p^2-p_{\mu}p_{\nu}) \\
   \end{equation}
   \begin{equation}
  P_{\mu\nu}(p,q) \stackrel{p,q\rightarrow0}{\longrightarrow}
 {1 \over 8\pi^2m} \epsilon_{\mu\nu\alpha\beta} q_{\alpha} p_{\beta}
   \end{equation}
We should mention that all the diagrams have been computed for QED rather
than QCD. This doesn't affect the comparison between lattice and continuum
at lowest order in perturbation theory and avoids the unnecessary burden of
carrying color factors all the way through the computation. In fact, once
the matrix element of a local fermionic current has been computed at lowest
order in an external QED field and expressed in gauge invariant terms, the
result can be generalized immediately to QCD (Color factors have to be added
but there is no need to compute explicitely the 3 or 4 external gluon lines
diagram which are required to insure gauge invariance in QCD).

\section{Lattice results and finite $ma$ correction factors}

We now want to compute the perturbative ``response functions'' on the lattice
in order to compare them with the continuum results presented above. The
computation will be done both for the Wilson action and for the 2-link
improved action of Hamber and Wu (see appendix for our conventions and
Feynman rules). As before, the actual calculations will be carried out
in the context of QED. Lattice perturbation theory is characterized in
general by the appearance of new Feynman diagrams with ``multi-photon''
vertices. In our case, the ``fish diagram'' (fig.2.c) appears
in addition to the two triangle diagrams (fig.2.a and b).
The lattice expression for diagrams of the type of figure 2.a is:
   \begin{equation}
   I^\Gamma_{\mu\nu}(p,q) =
   \int {d^4k \over (2\pi)^4} \mbox{Tr} \left[
   \Gamma S_F(k-q) \Gamma_\nu(k,-q) S_F(k) \Gamma_\mu(k,p) S_F(k+p)
   \right]
   \end{equation}
where the integral over $k$ extends from $-\pi$ to $+\pi$ in each direction
and the vertex $\Gamma_\mu$ and propagator $S_F$ are defined in the appendix.
The expression
for the diagram of fig.2.b can be obtained directly from (20) by the exchange
$p,\mu \leftrightarrow q,\nu$. Finally, the fish diagram fig.2.c gives the
integral
   \begin{equation}
   J^\Gamma_{\mu\nu}(p,q) =
   \int {d^4k \over (2\pi)^4} \mbox{Tr} \left[
   \Gamma S_F(k-q) \Gamma^{(2)}_{\mu\nu}(k+p,-p,-q) S_F(k+p)
   \right]
   \end{equation}
where $\Gamma^{(2)}_{\mu\nu}$ is defined in the appendix (eq.(37)).

Let's begin by describing the computation for the pseudoscalar current
which is probably the simplest of all. Note that in this case
($\Gamma = \gamma_5$) the trace in (21) vanishes so that only the integral
(20) has to be computed. Replacing $\Gamma$ by $\gamma_5$,
using the formulas (36) and (38) in the Appendix,
and taking the Dirac trace, we get:
\begin{eqnarray}
I^5_{\mu\nu}(p,q) &=&
  \int {d^4k \over (2\pi)^4} { 1 \over Z(k-q) Z(k) Z(k+p) } \nonumber \\
 & &\{4\epsilon_{\mu\nu\beta\delta} V_\mu(k,p) V_\nu(k,-q)
         M(k-q)S_\beta(k)S_\delta(k+p) \nonumber \\
 & & -4\epsilon_{\mu\nu\beta\delta} V_\mu(k+p) V_\nu(k,-q)
         M(k)S_\beta(k-q)S_\delta(k+p) \\
 & & +4\epsilon_{\mu\nu\beta\delta} V_\mu(k+p) V_\nu(k,-q)
         M(k+p)S_\beta(k-q)S_\delta(k)  \nonumber \\
 & & -4r\epsilon_{\mu\alpha\beta\delta} V_\mu(k,p) W_\nu(k,-q)
         S_\alpha(k-q)S_\beta(k)S_\delta(k+p) \nonumber \\
 & & -4r\epsilon_{\alpha\nu\beta\delta} W_\mu(k,p) V_\nu(k,-q)
         S_\alpha(k-q)S_\beta(k)S_\delta(k+p)  \} \nonumber
\end{eqnarray}

The integral over $d^4k$ can then be carried out numerically for the Wilson
or the improved action by using the explicit expression given in the appendix
for the various functions S,M,Z,V,W. However, since we are mostly interested
in the case of small momenta ($ p,q \rightarrow 0$), it is interesting to
go one step further in the analytical evaluation and carry out the Taylor
expansion of (22) up to second order in $p$ and $q$. This computation is
simplified by the fact that only the numerator needs to be expanded to
second order (it turns out that the terms coming from the expansion of the
denominator vanish because of the antisymmetry of the $\epsilon$ symbol).
The final result for Wilson fermions is ($ p,q \rightarrow 0$):
   \begin{equation}
   I^5_{\mu\nu}(p,q) =
   4\epsilon_{\mu\nu\alpha\beta} q_\alpha p_\beta
   \int { d^4k \over (2\pi)^4 } { 1 \over Z^3(k) }
   \left( \prod_{\rho=1}^4 \cos k_\rho \right)
   \left[ M(k)-r\sum_{\lambda} {\sin^2 k_\lambda \over \cos k_\lambda} \right]
   \end{equation}
And for the 2-link improved action, we have:
   \begin{equation}
   I^5_{\mu\nu}(p,q) =
   4\epsilon_{\mu\nu\alpha\beta} q_\alpha p_\beta
   \int { d^4k \over (2\pi)^4 } { 1 \over Z^3(k) }
   \left( \prod_{\rho=1}^4 C_\rho(k) \right)
   \left[ M(k)-r\sum_{\lambda} {S_\lambda(k) \over C_\lambda(k)}
   T_\lambda(k) \right]
   \end{equation}
where we have introduced:
   \begin{equation}
   C_\lambda (k) = {1\over3} (4\cos k_\lambda - \cos 2k_\lambda )
   \end{equation}
   \begin{equation}
   T_\lambda (k) = {1\over3} (4\sin k_\lambda - 2\sin2k_\lambda )
   \end{equation}
The integrals in (23) and (24) were evaluated numerically by using a
trapezoidal
rule on a non-uniform grid. The resolution was chosen to be maximum around
$k=0$ where the integrand peaks (especially when $ma$ is small). In practice,
we mostly used a set of nested $(4)^4$ grids where at each step, the resolution
is increased by a factor of 2 in the inner portion of the grid (see fig.3 for
an example). In order to control the discretization errors, we further varied
the resolution on each of the embedded grids separately and checked for
convergence. The nesting procedure is stopped when the contribution from
the inner grid becomes smaller than the overall discretization error. In
practice, 10 levels of nesting or less were always sufficient.
Once the numerical integrals have been computed, we can sum the contributions
from the diagrams of fig.2 and compare the result
with the continuum formula (19). In figure 4, we plot the ratio of
the lattice to the continuum results for the Wilson and improved action as
a function of $ma$ (at $r=1$). All curves agree at $ma=0$ (as they should), but
it is clear that elsewhere, Wilson fermions strongly underestimate the matrix
elements of the pseudoscalar current. The situation is significantly improved
in the case of the 2-link action.  These results are in agreement with those of
Hamber and Wu \cite{Wu84} and Smit and Vink \cite{Smit87}.
In tables 1 and 2, we collect the finite renormalization factors to
be used in lattice simulations of sea quark matrix elements (inverse of the
quantities plotted in fig. 4). We do so for the range of masses $0 < ma < 0.2$
relevant for current simulations and including the mass of the strange quark.
The precision on these renormalization factors is always better than $1\%$.
For comparison, we also include the corresponding factors for valence Wilson
quarks
($1+ma$). We would like to remind the reader that the numbers presented in
tables 1 and 2 represent only the $ma$ dependent part of the renormalization
factors, the usual perturbative renormalization factors have to be included
separetely.

Let's now consider the case of the axial current. We have to compute (20)
and (21)
with $\Gamma$ replaced by $i\gamma_\alpha \gamma_5$ [$i$ is added in order to
get a real answer]. Taking the trace directly in (20) would however lead to a
rather lengthy expression. We will therefore restrict our attention to the
case $q=-p$, which is sufficient for our purposes [zero momentum matrix
elements of $j_{\alpha5}$]. Note that for $q=-p$, the fish diagram for the
axial current [(21) with $\Gamma$ replaced by $i\gamma_\alpha \gamma_5$]
will vanish by the antisymmetry of the $\epsilon$ symbol.
For (20) we then obtain:
    \begin{eqnarray}
I^{\alpha5}_{\mu\nu}(p,-p)
&=& \int { d^4k \over (2\pi)^4 } { 1 \over Z^2(k) Z(k+p) } \nonumber \\
& & 4 \{  {1\over2} \epsilon_{\lambda\mu\nu\delta} V_\mu(k,p) V_\nu(k,p)
    [ 2 S_\delta(k) D(k,k+p) - S_\delta(k+p) Z(k) ] \nonumber \\
& & - 2\epsilon_{\lambda\mu\beta\delta} V_\mu(k,p) V_\nu(k,p)
     S_\beta(k+p) S_\delta(k) S_\nu(k) \\
& & - 2\epsilon_{\lambda\mu\beta\delta} V_\mu(k,p) W_\nu(k,p)
     rM(k) S_\beta(k+p) S_\delta(k) \nonumber \\
& & - (\mu \leftrightarrow \nu) \} \nonumber
    \end{eqnarray}
where we have introduced:
    \begin{equation}
D(k,k+p)=\sum_\alpha S_\alpha(k) S_\alpha(k+p) + M(k)M(k+p)
    \end{equation}
As in the pseudoscalar current case, one could in principle expand the integral
in powers of p. However this is rather difficult in the present case. One
can show for example that the lowest non-trivial order is $O(p^3)$. Order
$0$ and $2$ vanish by symmetrical integration and order one is a total
divergence. Therefore,
the expansion appears rather complicated and not really interesting given
that the final integration will be computed numerically anyway. We therefore
opted for the numerical computation of (27) at various values of $p$ and an
extrapolation to $p=0$. The numerical integration proceeds as before except
that now, there are 2 points where the integrand peaks ($k=0$ and $k=-p$).
We take this into account by refining the grid separately around these 2
points. Our integration routine generates the grids and does all the
bookkeeping automatically. The procedure is in fact rather general and is
in a sense a restriction of the adaptative quadrature method to the case
where the position of ``troublesome'' points is known in advance.
Discretization errors are controlled as before by doubling the resolution
on each of the nested grids separately. For small momentum, the results of
the numerical integration can be compared with the continuum formula (17).
The inverse ratio $1/\kappa_A$ is plotted in  fig.5 and the
axial correction factor $\kappa_A$ presented in tables 1 and 2. The results are
qualitatively similar to those obtained for $j_5$ except that the lattice
artifacts are smaller for the axial than for the pseudoscalar current.
Finally, it is also interesting to look at the situation at finite $pa$ as
this might become useful in a more thorough analysis of lattice artifacts.
That is, although we would hope that a numerical simulation is dominated by
gluonic fields of small $pa$, we have no reason to assume this a priori and
it might be important to look at ``rough'' fields too.
We would therefore like to fix $ma$ and compare the momentum
dependence of the lattice integrals with the continuum formula (10). In fig.6,
we present the result of such a comparison with $ma$ fixed at $0.2$
(the continuum solid curve corresponds to $A(p^2/m^2)$ as given by (13)
and the dots are results obtained on a grid of lattice momenta).
We have limited the comparison to the range $|pa|\leq2\pi$ accessible on the
lattice.

Finally, let's consider the scalar current. Note that in this case,
the ``fish'' diagram (fig.2.c) is non-zero and has to be included as well.
Only the sum of all 3 diagrams in fig.2 gives a gauge invariant result.
In fact, we checked that the final results obtained in this way are
proportional to a lattice transverse tensor which reduces to
$(\delta_{\mu\nu} - p_\mu p_\nu / p^2)$ in the low momentum limit. In our
most extensive comparisons between lattice and continuum, however, we didn't
compute the full tensor but rather focused our attention on a single
off-diagonal element. The fish diagram, being proportional to
$\delta_{\mu\nu}$, doesn't appear in this case and this helps in further
reducing the computer time requirements. The expression for the triangle
diagram (20) with $\Gamma=1$ and $q=-p$ is:

\begin{eqnarray}
I_{\mu\nu}(p,-p)
&=&\int { d^4k \over (2\pi)^4 } { 1 \over Z^2(k) Z(k+p) } \nonumber \\
& &\{
   V_\mu(k,p) V_\nu(k,p) 8 M(k) \left[ S_\mu(k+p) S_\nu(k)
                                + S_\mu(k) S_\nu(k+p) \right] \nonumber \\
& & +4r V_\mu(k,p) W_\nu(k,p) \left[
          2 M(k) F_\mu(k,k+p) - S_\mu(k+p) Z(k) \right] \nonumber \\
& & +4r W_\mu(k,p) V_\nu(k,p) \left[
          2 M(k) F_\nu(k,k+p) - S_\nu(k+p) Z(k) \right] \\
& & -4r^2 W_\mu(k,p) W_\nu(k,p) \left[
          2 M(k) E(k,k+p) + M(k+p) Z(k) \right] \nonumber \\
& & -4 \delta_{\mu\nu} \left[ V_\mu(k,p) \right]^2 \left[
          2 M(k) D(k,k+p) - M(k+p) Z(k) \right] \} \nonumber
\end{eqnarray}
where we have introduced:
   \begin{equation}
E(k,k+p) = \sum_\alpha S_\alpha(k) S_\alpha(k+p) - M(k) M(k+p)
   \end{equation}
   \begin{equation}
F_\mu(k,k+p) = S_\mu(k) M(k+p) + S_\mu(k+p) M(k)
   \end{equation}

The numerical integration of (29) proceeds as in the case of the axial current.
The computations are made for various values of p and extrapolated down to
$p=0$. Comparing the small momentum lattice results with the continuum formula
(18) gives the renormalization factor $\kappa_S(ma)$ defined in (8). The
inverse
of this factor is plotted in fig.7. It appears that when $ma$ is non-zero and
moderate, Wilson fermions overestimate the continuum result. The deviation is
significant, being almost $30\%$ at $ma=0.1$. (The renormalization factor is
tabulated in table 1). At large values of $ma$ on the other hand, the lattice
would underestimate the continuum result. The Hamber-Wu action provides a
moderate improvement for small values of $ma$. However for $ma \ge 0.1$,
the situation is as bad or even worse than for Wilson fermions (The correction
factors for the improved action are given in table 2).
We also made a comparison of lattice and continuum results
at finite momentum. By fixing $ma=0.2$, we obtain the results presented in
figure 8. The continuum solid curve corresponds to the function $S(p^2/m^2)$
as given by (14). Note that up to an overall factor, this is precisely the
comparison between the functions L and C [eqs. (6) and (7)] introduced in
section 3 [eq. (11)
provides the exact correspondence]. From the curves of figure 8, it appears
that for background fields of non-zero momentum, Wilson fermions overestimate
matrix elements of the scalar current even more than in the zero momentum
limit. It is interesting to contrast this with what happens in the case of
naive fermions (There of course, one divides the lattice results by 16 before
making the comparison with the continuum). The results are presented in
fig.9 and turn out to be much better than for the Wilson case. It is not
entirely clear what the origin of the problem is with Wilson fermions.
They probably essentially come from the form of the quark-gluon vertex
[i.e. eq.(36)].
Gluons do not only couple to the quarks through $\gamma_\mu$ but also directly
through a ``mass term'' $rW_\mu$. In fact, one can check that by artificially
truncating the vertex (taking $r=0$ in (36)) in the Wilson fermion calculation
of the
triangle diagram and leaving the propagators unchanged, one obtains a result
very similar  to fig.9. The result of fig.8 indicates that one has to be
very careful when using Wilson fermions to measure matrix elements of the
scalar current on the lattice. Simulation results might be even more
overestimated
than we would think from the low momentum analysis given above and leading to
the scalar correction factors given in table 1. This will have to be
analyzed in more detailed investigations.

\section{Examples of phenomenological applications}

The results presented above indicate that the finite $ma$ correction factors
for sea quarks are in general rather large and will therefore strongly
influence the interpretation of simulation results. To illustrate this,
we now discuss their application to two interesting problems \cite{Lat94b}.
The first is the computation of the mass shift of the $\eta^\prime$.
In a quenched simulation, it can be estimated from a measurement of the
coupling between the would be Goldstone bosons. The computation involves
the ratio of a 2-loop (fig.1.d) to 1-loop (fig.1.c) amplitude \cite{Kura94} .
As a consequence of the above discussion, we expect that they
will have different renormalization factors: The 1-loop part behaves as a
``valence'' quark and will therefore pick-up a ($1+ma$) correction for
each current whereas the 2-loop diagram corresponds to a ``sea'' quark
situation, in which case, the pseudoscalar factor of table 1 is relevant.
In figure 10, we compare the mass of the $\eta^\prime$ after correction with
the raw lattice data \cite{Kura94}.
 The trends are very different: after correction,
the mass shift of the $\eta^\prime$ is essentially independent of the mass
of the quark up to the strange quark mass with a chiral limit that is in much
better agreement with the estimate obtained from the Witten-Veneziano formula
(We assume that the drop in the renormalized data at the lowest quark mass
can be attributed to finite size effects or to the zero-mode
shift effect \cite{Smit87}). The agreement between the renormalized data and
the Witten-Veneziano estimate $M_0^2=2N_f\chi/f_\pi^2$ doesn't come as a
surprise,  since the correction factor that we are advocating
here is the same as the one used by Smit and Vink \cite{Smit87} to compute
 the topological susceptibility by the fermionic method [See eq.(5)].
It is just a matter of self-consistency that the correction factor appearing
in (5) be used as well in the computation of the 2-loop amplitude [Fig.1.d]
as a function of time separation. Otherwise, the Witten-Veneziano relation
could not be recovered on the lattice. Note also that the (quark) mass
independence of the coupling between the would be Goldstone bosons is a basic
assumption in the Witten-Veneziano mass formula. It is nice to be able to
verify it on the lattice.  As a second example,
we mention the contribution of the strange quark to the mass of the nucleon.
If we replace the ($1+ma$) correction factor used in \cite{Fuku94} with the
sea quark scalar correction factor from table 1, we find that the strange
quark contribution to the mass of the nucleon comes down,
from $30\%$ to $21\%$ - a change of $30\%$.

\section{Conclusions}

We have computed the $ma$ dependent part of lattice renormalization factors
for sea quark matrix elements. The results depend strongly on the Lorentz
structure of the current under consideration.
This is unlike the valence quark situation where
in first approximation, a single $ma$ dependent correction factor can
be used for all currents. We found for example that for Wilson fermions,
sea quark matrix elements of the axial and pseudoscalar current are
underestimated whereas those of the scalar current are overestimated.
The perturbative method that we have used for computing the finite $ma$
correction factors is general and could be used for other local currents,
as well as for point-split currents. The $ma$ dependent corrections for sea
quarks are rather large and generally dominate the remaining part of the
renormalization factor ($2\tilde\kappa[1+O(\alpha_V)]$), which of course
underlies their phenomenological importance. In section 6, we applied our
renormalization factors to the results of recent lattice simulations
concerning the mass shift of the $\eta^\prime$ \cite{Kura94} and the strange
quark content of the nucleon \cite{Fuku94}. In both cases, it was found that
the finite $ma$ correction had a large impact and helped in bringing the
simulation result closer to phenomenological estimates.

There is also much interest in trying to reduce the magnitude of these
correction factors by using modified actions. Our results for the axial and
pseudoscalar currents for example give a strong motivation for using the
2-link improved action of Hamber and Wu in these cases. For the scalar
current, on the other hand, this kind of improvement doesn't seem to help
much and it is probably more interesting
to consider other alternatives, such as staggered fermions for example or
working with naive fermions and dividing the result by 16.
Finally, we would like to recall that the renormalization factors
quoted in tables 1 and 2 have been obtained
assuming that the lattice simulation is dominated by low momentum gauge
fields. What happens exactly in more general situations will have to be
investigated
in later studies. In the final analysis, it seems to us that the ideas
presented here would be best tested by doing simulations using different
actions (Wilson versus improved for example or Wilson at different values
of r), applying the appropriate correction factor in each case and then
checking the overall compatibility of the physical results.

\section{Appendix}

The action for Wilson and two-link inproved fermions is usually defined
as \cite{Wu84}:
\begin{eqnarray}
A_{\mbox{F}} &=& \kappa \sum_{n,\mu} \left[
   \bar\psi_n (r-\gamma_\mu)U_{n,\mu} \psi_{n+\mu} +
   \bar\psi_{n+\mu} (r+\gamma_\mu)U^\dagger_{n,\mu}\psi_n \right]
   \nonumber \\
&+& \sum_{n,\mu} \left[ \bar\psi_n (C-D\gamma_\mu) U_{n,\mu} U_{n+\mu,\mu}
   \psi_{n+2\mu} + \bar\psi_{n+2\mu} (C+D\gamma_\mu) U^\dagger_{n+\mu,\mu}
   U^\dagger_{n,\mu}\psi_n \right]
   \nonumber \\
&-& \sum_n \bar\psi_n\psi_n
\end{eqnarray}
For the Wilson action, $C=D=0$; whereas for the improved action, $C$ and $D$
are chosen to cancel all terms of order $(pa)^2$ and $(pa)^3$ in the
expansion of the action. One then finds:
   \begin{equation}
   C={ -\kappa r \over 4 } \quad\quad\quad D={ -\kappa \over 8 }
   \end{equation}
It has to be remarked, however, that with this definition, the $0^{th}$ order
term and the terms linear in $(pa)$ are modified by the improvement procedure.
This is usually corrected at the end of the computation by introducing a
wave function renormalization and by changing the definition of the quark
mass in terms of $\kappa$ ( $ma={2 \over 3\kappa} -4r$ ). Here however, we
will choose the alternative solution of making the correction from the start.
That is, in the {\it improved} case, we will multiply the action (32) by $4/3$.
This will insure that our Feynman rules have the correct continuum
(=low momentum) limit. In this way, the relation between $ma$ and $\kappa$
will also be the same for the two actions:
   \begin{equation}
   ma = { 1 \over 2\kappa } - 4r
   \end{equation}
Finally, in the same spirit of keeping things as close as possible to the
continuum formulation, we absorb the factor of $\sqrt{2\kappa}$ in the
definition of $\psi$. With these remarks, we can now introduce our Feynman
rules. Our definitions are the same as in \cite{Bern87}. We write the inverse
fermion propagator in the form:
   \begin{equation}
   S^{-1}_F(p) = \sum_{\alpha} i\gamma_\alpha S_\alpha(p) + M(p)
   \end{equation}
the 1-photon vertex (fig.11.a) in the form:
   \begin{equation}
   \Gamma_\mu(k,q)= i\gamma_\mu V_\mu(k,q) + r W_\mu(k,q)
   \end{equation}
and the 2-photon vertex (fig.11.b) in the form:
   \begin{equation}
   \Gamma^{(2)}_{\mu\nu}(k,q,p)= \delta_{\mu\nu} \left(
   i\gamma_\mu V^{(2)}_\mu(k,q,p) - r W^{(2)}_\mu(k,q,p) \right)
   \end{equation}
{}From (35), we find for the fermion propagator:
   \begin{equation}
   S_F(p) = {-i\sum_{\alpha} \gamma_\alpha S_\alpha(p) + M(p) \over Z(p)}
   \end{equation}
where we have defined
   \begin{equation}
   Z(p)=\sum_\beta S^2_\beta(p)+M^2(p)
   \end{equation}
We then have for Wilson fermions:
   \begin{equation}
              S_\alpha(p) = \sin p_\alpha
   \end{equation}
   \begin{equation}
              M(p)=m+4r-r\sum_\beta\cos p_\beta
   \end{equation}
   \begin{equation}
              V_\mu(k,q)=\cos(k_\mu + {q_\mu \over 2})
   \end{equation}
   \begin{equation}
              W_\mu(k,q)=\sin(k_\mu + {q_\mu \over 2})
   \end{equation}
   \begin{equation}
        V^{(2)}_\mu(k,q,p)=\sin(k_\mu + {q_\mu \over 2} + {p_\mu \over 2})
   \end{equation}
   \begin{equation}
        W^{(2)}_\mu(k,q,p)=\cos(k_\mu + {q_\mu \over 2} + {p_\mu \over 2})
   \end{equation}
and for the improved action:
   \begin{equation}
           S_\alpha(p) = {1\over6} (8\sin p_\alpha - \sin 2 p_\alpha)
   \end{equation}
   \begin{equation}
    M(p) = m+4r- {r\over3}\sum_\beta (4\cos p_\beta - \cos 2p_\beta )
   \end{equation}
   \begin{equation}
           V_\mu(k,q)= {1\over3} \left[ 4 \cos(k_\mu+ {q_\mu\over2})
           -cos({q_\mu\over2}) cos(2k_\mu+q_\mu) \right]
   \end{equation}
   \begin{equation}
          W_\mu(k,q)= {1\over3} \left[ 4 \cos(k_\mu+ {q_\mu\over2})
         -2cos({q_\mu\over2}) sin(2k_\mu+q_\mu) \right]
   \end{equation}
   \begin{equation}
        V^{(2)}_\mu(k,q,p)= {1\over3} \left[
        4 \sin(k_\mu + {q_\mu \over 2} + {p_\mu \over 2})
       -2 \cos({q_\mu \over 2}) \cos({p_\mu \over 2})
          \sin( 2 k_\mu + q_\mu + p_\mu ) \right]
   \end{equation}
   \begin{equation}
        W^{(2)}_\mu(k,q,p)= {1\over3} \left[
        4 \cos(k_\mu + {q_\mu \over 2} + {p_\mu \over 2})
       -4 \cos({q_\mu \over 2}) \cos({p_\mu \over 2})
          \cos( 2 k_\mu + q_\mu + p_\mu ) \right]
   \end{equation}
In writing these Feynman rules, we have omitted all the factors of the lattice
spacing a. These however are obvious on dimensional grounds and can easily be
reinstated at the end of the computation if necessary. Finally, the continuum
Feynman rules are obtained straightforwardly by taking the low momentum limit
of the above formulas. The result is:
   \begin{equation}
S_\alpha(p)=p_\alpha \quad\quad M(p)=m \quad\quad V_\mu = 1 \quad\quad
W_\mu=V^{(2)}_\mu =W^{(2)}_\mu= \cdots = 0
   \end{equation}

\newpage
{\Large {\bf Table Caption}}
\begin{enumerate}
\item $ma$ dependence of local ``sea'' currents renormalization factors
      for Wilson fer\-mions (r=1).
\item Same as table 1 for Hamber-Wu fermions.
\end{enumerate}
\vspace{1cm}

{\Large {\bf Figure Caption}}
\begin{enumerate}
\item Quark skeleton diagrams associated with various measurements:
      (a) Connected and (b) disconnected contributions to a matrix
      element in the nucleon. (c) Connected and (d) disconnected
      contributions to the propagator of a flavor-singlet meson.
\item Leading order diagrams contributing to the matrix element of
      a current in an external field (Diagram (c) appears only on
      the lattice).
\item Example of a grid used for numerical integration.
\item Pseudoscalar current response to a given external field. The ratio
      of the lattice and continuum responses is plotted versus $ma$ for
      two choices of the lattice action.
\item Same as fig.4 for the axial current.
\item Dependence of the axial current response function on the momentum of
      the external gauge field ($ma$ is fixed at $0.2$ and $pa$ varies
      between $0$ and $2\pi$). The continuum response function (solid line)
      is included for comparison.
\item Same as fig.4 for the scalar current.
\item Same as fig.6 for the scalar current.
\item Same as fig.8 for naive fermions (after division by $16$).
\item $\eta^\prime$ mass shift vs quark mass after renormalization (squares)
      and before (plusses). Also included are the chiral extrapolation of the
      raw data (diamond) and the estimate from the Witten-Veneziano formula
      (circle), using the topological susceptibility $\chi$ obtained in
      \cite{Kura94}.
\item One and two photon vertices.
\end{enumerate}

\newpage
\newcommand{\x}{\hspace{1em}}
\newcommand{\xx}{\hspace{2em}}
\begin{tabular}{|c|c|c|c|c|}
\hline
& valence  & \multicolumn{3}{c|}{sea quarks} \\
\cline{3-5}
\hbox{\x ma \x} & quarks & pseudoscalar ($\kappa_P$) &
\hbox{\x axial ($\kappa_A$)\x}
 & \hbox{\x scalar ($\kappa_S$)\x} \\
\hline
 0.00 & 1.00 & 1.000 & 1.000 & 1.000 \\
 0.02 & 1.02 & 1.199 & 1.045 & 0.865 \\
 0.04 & 1.04 & 1.373 & 1.093 & 0.816 \\
 0.06 & 1.06 & 1.550 & 1.146 & 0.795 \\
 0.08 & 1.08 & 1.732 & 1.204 & 0.787 \\
 0.10 & 1.10 & 1.922 & 1.265 & 0.787 \\
 0.12 & 1.12 & 2.121 & 1.329 & 0.793 \\
 0.14 & 1.14 & 2.331 & 1.397 & 0.803 \\
 0.16 & 1.16 & 2.552 & 1.467 & 0.816 \\
 0.18 & 1.18 & 2.784 & 1.542 & 0.832 \\
 0.20 & 1.20 & 3.029 & 1.619 & 0.849 \\
\hline
\end{tabular}

\vspace{1cm}
\center{Table 1}
\vspace{2cm}

\begin{tabular}{|c|c|c|c|}
\hline
& \multicolumn{3}{c|}{sea quarks} \\
\cline{2-4}
\hbox{\x ma \x} & pseudoscalar ($\kappa_P$)
& \hbox{\xx axial ($\kappa_A$)\xx}
 & \hbox{\xx scalar ($\kappa_S$)\xx} \\
\hline
 0.00 & 1.000 & 1.000 & 1.000 \\
 0.02 & 1.038 & 1.000 & 0.946 \\
 0.04 & 1.078 & 1.001 & 0.894 \\
 0.06 & 1.120 & 1.004 & 0.852 \\
 0.08 & 1.165 & 1.007 & 0.816 \\
 0.10 & 1.212 & 1.012 & 0.784 \\
 0.12 & 1.261 & 1.018 & 0.758 \\
 0.14 & 1.312 & 1.026 & 0.735 \\
 0.16 & 1.366 & 1.034 & 0.715 \\
 0.18 & 1.422 & 1.044 & 0.697 \\
 0.20 & 1.480 & 1.055 & 0.682 \\
\hline
\end{tabular}
\vspace{1cm}
\center{Table2}


\begin{thebibliography}{99}
\bibitem{Liu94}  K.-F. Liu and S.-J. Dong, Phys. Rev. Lett. 72, 1790 (1994)
\bibitem{Dong94} S.-J. Dong and K.-F. Liu, Phys. Lett. B328, 130 (1994)
\bibitem{Lat94a}  S.-J. Dong and K.-F. Liu, $\pi N \sigma$ Term and Quark Spin
                 Content of the Nucleon, UK/94-07 (hep-lat/9412059)
\bibitem{Kura94} Y. Kuramashi, M. Fukugita, H. Mino, M. Okawa and A. Ukawa,
                 Phys. Rev. Lett. 72, 3448 (1994)
\bibitem{Fuku94} M. Fukugita, Y. Kuramashi, M. Okawa and A. Ukawa, Pion-Nucleon
                 sigma term in lattice QCD, UTHEP-281, hep-lat/9408002
\bibitem{Marti83} G. Martinelli and Y.-C. Zhang, Phys. Lett. B123, 433(1983);
                  B125,77(1983)
\bibitem{Lepa92} G.P. Lepage, Nucl. Phys. B (Proc. Suppl.) 26 (1992) 45;
                 A.S. Kronfeld, Nucl. Phys. B (Proc.Suppl.) 30 (1993) 445
\bibitem{Lepa93} G.P. Lepage and P. Mackenzie, Phys. Rev. D48 (1993) 2250
\bibitem{Wu84}   H.W.Hamber and C.M.Wu, Phys. Lett. 136B, 255 (1984)
\bibitem{Smit81} L.H. Karsten and J.Smit, Nucl. Phys. B183 (1981) 103
\bibitem{Smit87} J. Smit and J.C. Vink, Nucl. Phys. B286 (1987) 485,
                 B307 (1988) 549; Phys. Lett. B194, 433 (1987)
\bibitem{SVZ}    M.A. Shifman, A.I. Vainshtein and V.I. Zakharov,
                 Nucl. Phys. B147 (1979) 385, 448, 519
\bibitem{Lat94b}  J.-F. Laga\"{e} and K.-F. Liu,
                 Finite $ma$ corrections for sea quark matrix elements,
                 UK/94-05, hep-lat/9412023, To appear in the proceedings
                 of the LATTICE'94 conference, Nucl. Phys. B (Proc. Suppl.
                 section)
\bibitem{Bern87} C. Bernard, A. Soni and T. Draper, Phys. Rev. D36, 3224 (1987)
\end{thebibliography}
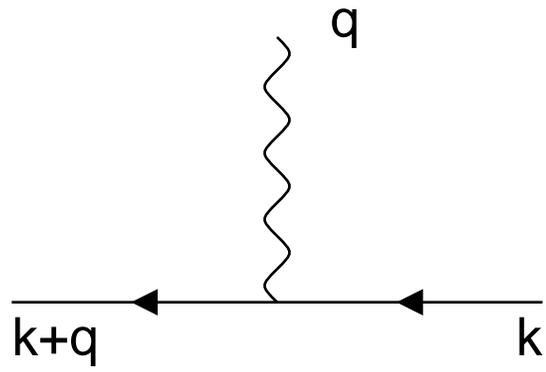
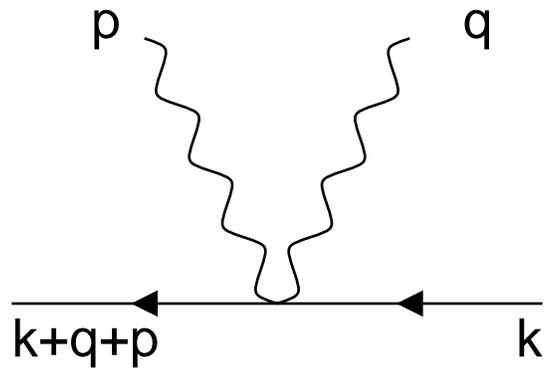
\end{document}